\documentclass[twocolumn,pre,superscriptaddress,floatfix,letterpaper]{revtex4}
\usepackage{amsmath,amssymb,amsfonts}
\usepackage{graphicx}
\usepackage[urlcolor=blue, hyperindex, colorlinks, bookmarks=true]{hyperref}
\usepackage{natbib}
\usepackage{times}
\usepackage{upgreek}

\newcommand{\ket}[1]{|#1)}
\newcommand{\bra}[1]{(#1|}
\newcommand{\bket}[2]{(#1|#2)}
\newcommand{\boket}[3]{(#1|#2|#3)}

\newcommand{\ts}[1]{\alpha_{#1}}
\newcommand{\wdr}{\omega_{d}}

\newcommand{\mT}{\mathsf{T}}

\begin{document} 

\title{Imaging Photon Lattice States by Scanning Defect Microscopy}
\author{D.\ L.\ Underwood}
\altaffiliation[Current affiliation: ]{HRL Laboratories, LLC, 3011 Malibu Rd
Malibu, CA 90265, USA}
\affiliation{Department of Electrical Engineering, Princeton University, Princeton, New Jersey 08544, USA}
\author{W.\ E.\ Shanks}
\altaffiliation[Current affiliation: ]{IBM T.\ J.\ Watson Research Center, 1101 Kitchawan Rd, Yorktown Heights, NY 10598, USA}
\affiliation{Department of Electrical Engineering, Princeton University, Princeton, New Jersey 08544, USA}
\author{Andy C.\ Y.\ Li}
\affiliation{Department of Physics and Astronomy, Northwestern University, Evanston, Illinois 60208, USA}
\author{Lamia Ateshian}
\affiliation{Department of Electrical Engineering, Princeton University, Princeton, New Jersey 08544, USA}
\author{ Jens Koch}
\affiliation{Department of Physics and Astronomy, Northwestern University, Evanston, Illinois 60208, USA}
\author{ A.\ A.\ Houck}
\affiliation{Department of Electrical Engineering, Princeton University, Princeton, New Jersey 08544, USA}%

\date{October 28, 2015}%

\begin{abstract}
Microwave photons inside lattices of coupled resonators and superconducting qubits can exhibit surprising matter-like behavior. Realizing such open-system quantum simulators presents an experimental challenge and requires new tools and measurement techniques. Here, we introduce Scanning Defect Microscopy as one such tool and illustrate its use in mapping the normal-mode structure of microwave photons inside a 49-site Kagome lattice of coplanar waveguide resonators. Scanning is accomplished by moving a probe equipped with a sapphire tip across the lattice. This locally perturbs resonator frequencies and induces shifts of the lattice resonance frequencies which we determine by measuring the transmission spectrum. From the magnitude of mode shifts we can reconstruct photon field amplitudes at each lattice site and thus create spatial images of the photon-lattice normal modes.
\end{abstract}
\maketitle 

\section{Introduction}
Impressive experimental advances over the last two decades have turned the idea of quantum simulation \cite{Feynman1982} into a reality \cite{Jaksch1998,Greiner2002,Bloch2005,Buluta2009,Bloch2012,Blatt2012,Aspuru-Guzik2012}. The use of minutely controlled quantum systems with systematic tunability of parameters such as interaction strength and particle density, has opened the door for the experimental study of complex many-body problems. Today, a variety of physical implementations of analog quantum simulators exists, and their primary focus has been the realization of models in equilibrium, often close to zero temperature -- the paradigmatic example being the study of the Bose-Hubbard model with ultracold atoms inside an optical lattice \cite{Jaksch1998,Greiner2002}.

Proposals for photon-based quantum simulation \cite{Greentree2006,Hartmann2006,Angelakis2007} have recently received a flurry of interest and stimulated a host of theoretical studies (see Refs.\ \onlinecite{Tomadin2010,Houck2012a,Schmidt2013} for reviews). The lack of number conservation distinguishes photons from other bosonic systems and renders photonic quantum simulators ideal candidates for studying many-body physics under well-controlled non-equilibrium conditions.   One suggested physical realization of photon-based quantum simulation consists of microwave photons inside large networks of superconducting resonators and circuits. This architecture is particularly promising due to the significant experimental progress in the field of circuit quantum electrodynamics (cQED) in which superconducting qubits are coupled to superconducting resonators \cite{Wallraff2004,Blais2004,Schoelkopf2008}. 

\begin{figure}
\centering
\includegraphics[width=0.95\columnwidth]{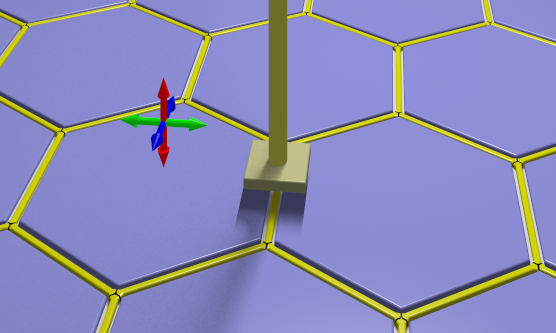}
\caption{Scanning Defect Microscopy of a photon lattice. A dielectric probe is scanned across a  photon lattice of coplanar waveguide resonators. The dielectric locally perturbs the electric field and shifts the frequency of a target resonator, creating a lattice defect. The transmission of photons through the lattice systematically depends on the defect strength and can be used to image the  photon occupancy on individual lattice sites.}
\label{device-art}
\end{figure}

The success of cQED systems stems from multiple factors, including the ability to readily reach the strong-coupling regime \cite{Wallraff2004,Bishop2008}, as well as the fact that cQED devices can be fabricated using standard lithography techniques. Most cQED research has primarily been motivated by the quest to implement a quantum computer. Some theoretical studies in the recent past have focused, instead, on quantum simulation, and predict that photons within large lattices of coupled cQED systems can exhibit striking matter-like behavior \cite{Carusotto2009,Kiffner2010,Hartmann2010}.

Very recently, the first experiment with a cQED-based quantum simulator has demonstrated a photon-number dependent crossover from a delocalized regime to a localized regime \cite{Raftery2013}. Despite the small lattice size of only two sites in that case, the exponential proliferation of Hilbert space dimension with increasing photon number quickly approaches  the computational limits of a classical computer, rendering the system a quantum simulator in the rigorous sense \cite{Cirac2012}.

Our work presented here focuses on the understanding and experimental study of larger photon-based lattices for quantum simulation.
Earlier efforts have already shown that microwave resonator lattices with very low disorder can be fabricated reliably \cite{Underwood2012}.  However, measuring and probing many-body states in such lattices still remains a significant experimental challenge. Large lattices form a complex network of planar microwave circuitry, but could, in a strict 2d setting, only be accessed via drives and detection probing the edge of the lattice at predetermined sites. While providing information about modes with non-zero amplitudes on edge sites, such measurements only give limited and indirect information about the bulk of interior lattice sites. 

To overcome these limitations,  here we introduce Scanning Defect Microscopy -- a novel  scanning-probe imaging technique applicable to coupled resonator and cQED arrays [Fig.\ \ref{device-art}]. Scanning Defect Microscopy makes it possible to gather local information about photon occupancies in such lattices. For instance, it allows us to locally image the normal modes in a microwave resonator array. In our experiment, we acquire these images by monitoring variations in microwave transmission when selectively altering the photon occupancy in one resonator. This is accomplished by positioning a small piece of dielectric precisely above the surface of a targeted resonator inside the lattice. The close proximity of the dielectric shifts the local resonator frequency and, thus, creates a lattice defect whose size can be tuned by controlling the vertical distance between the dielectric and the surface of the  resonator. Scanning the dielectric probe across the lattice and analyzing the systematic changes in the transmission spectrum due to the lattice defect reveal local information which is used, in our  example, to image the normal-mode photon occupancies across the resonator lattice. 

We note that similar methods, referred to as bead-pull experiments, are commonly used to characterize higher-order modes in large RF cavities for accelerators \cite{Maier1952,Bhat1993}. 
The conceptual idea underlying our scanning-probe scheme is also quite similar to scanning gate microscopy \cite{Topinka2000,Crook2000,Topinka2001} which images the electron flow in 2d electron-gas nanostructures. In scanning gate microscopy, image construction is based on monitoring the electric conductance when selectively depleting a small region of the electron gas by a scanning electrostatic gate. 

In the following, we present the fundamental basics of Scanning Defect Microscopy, and illustrate its use in obtaining images of the normal modes inside a photonic Kagome lattice of microwave resonators.

\section{Normal-Mode Imaging: Model\label{sec:model}}
\begin{figure*}
	\centering
	\includegraphics[width=0.95\textwidth]{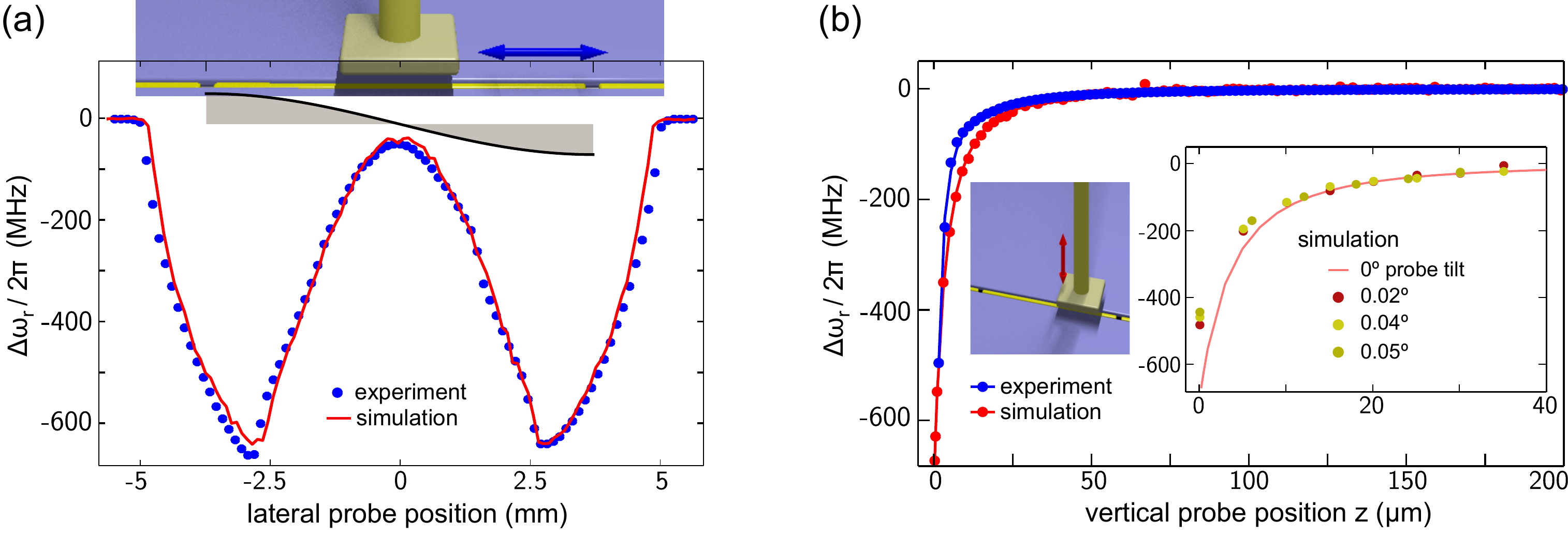}
	\caption{Verification of probe performance and calibration for single resonator. (a) 
		Measurement data of the resonator frequency shift as a function of probe position along the resonator agree well with predictions from finite-element simulation (using a $2\,\mu$m probe height). In the experiment, the probe is centered over and placed into mechanical contact with the resonator for each data point.  The frequency shifts follow the field distribution of the $\lambda/2$ mode, with shift maxima for probe positions close to the resonator ends where the field is strongest, and a shift minimum at the resonator midpoint. (b)  The calibration of the resonator frequency shift (defect size) with respect to vertical distance of the probe from the resonator is performed at the lateral position of maximum shift. After adjusting for an offset in $z$ direction, finite-element simulation and measurement data agree over most of the vertical distance range. Deviations observed at the smallest probe heights may be due to a small tilt of the dielectric probe with a misalignment angle as small as $0.02^\circ$, as indicated by simulations shown in the inset.}
	\label{YZscan}
\end{figure*}

To explain the underpinning for imaging of normal modes with Scanning Defect Microscopy, we consider the heterodyne transmission through a driven-damped resonator lattice as described by the Lindblad master equation  or input-output theory formalism \cite{Walls1995}. The lattice is coherently driven through one input port, attached to a particular edge site $n_\text{in}$ of the lattice. The transmission amplitude is measured through an output port, and  is directly proportional to the coherent-state amplitude $\langle a_\text{out}\rangle$ at the corresponding edge site $n_\text{out}$. The equation relating the coherent-state amplitude $\alpha_n=\langle\alpha_n\rangle$ at any lattice site $n$ to the drive amplitude $\epsilon$ is \cite{Nunnenkamp2011}
\begin{equation}
\left( \omega_{r} -\wdr- i \tfrac{1}{2}\kappa \right) \ts{n}+ t\sum_{\langle n,m \rangle} \ts{m} = -\epsilon \, \delta_{n,n_\text{in}},
\end{equation}
where $\omega_{r}$ and $\omega_d$ denote the bare resonator and drive frequency, $t$ the strength of photon hopping between neighboring resonators, and $\kappa$ the photon loss rate. 

The above equation is solved easily by matrix inversion
\begin{equation}
\ket{\alpha}= -\left[ ( \omega_{r}-\wdr -i\tfrac{1}{2}\kappa)\openone + t\,\mT   \right]^{-1}\ket{\epsilon},\label{amp-eq}
\end{equation}
where the real-symmetric matrix $\mT$ is the adjacency matrix of the lattice. For mere convenience, we use vector notation where $\ket{\alpha}$ collects the coherent-state amplitudes on each lattice site, $\bket{n}{\alpha}=\alpha_n$ and $\ket{\epsilon}$  encodes the drive on each site -- here acting only on the input port, $\bket{n}{\epsilon} = \epsilon\,\delta_{n,n_\text{in}}$. Transmission resonances naturally occur at the normal-mode frequencies $\Omega_\mu$ of the lattice, obtained from the usual normal-mode eigenvalue equation $(\omega_r\openone + t\,\mT)\ket{\mu} = \Omega_\mu \ket{\mu}$. As a result, the coherent-state amplitude at the output site takes the simple form 
\begin{equation}
\alpha_\text{out} =  -\sum_{\mu}\bket{n_\text{out}}{\mu}\frac{1}{(\Omega_\mu-\omega_d)+i\tfrac{1}{2}\kappa}\bket{\mu}{n_\text{in}}\, \epsilon.
\end{equation}
Accordingly, each normal mode with nonzero mode amplitude at both input and output port will produce a transmission peak when the drive frequency coincides with the normal-mode frequency. (This assumes frequency spacing between modes to be larger than $\kappa$, which is appropriate for the lattice investigated here.) 

Once a certain normal mode $\mu$ is on resonance, its normal-mode weight $W_{\mu n}=|\bket{n}{\mu}|^2$  on each lattice site $n$ can be extracted by introducing a small shift of the bare cavity frequency $\Delta \omega_n$. This lattice defect induces a frequency shift $\Delta\Omega_{\mu|n}$ in the normal-mode frequency. The connection between the small normal-mode frequency shifts and the mode weights is established by simple perturbation theory. We write the new normal-mode matrix including the defect as $\mathsf{H}_n=\omega_r\openone+ \Delta\omega_n\mathsf{D}_n+ t\,\mT $, where $\mathsf{D}_n = \ket{n}\bra{n}$ selects the site of the defect. Applying perturbation theory with respect to the defect $\Delta\omega_n\mathsf{D}_n$, we immediately obtain the leading-order frequency shift for normal mode $\mu$ caused by a defect on site $n$:
\begin{equation}
\Delta \Omega_{\mu|n}  = \Delta\omega_n\boket{\mu}{\mathsf{D}_n}{\mu}
 = \Delta\omega_n\, W_{\mu n}.
\end{equation}
 The weight of the (undisturbed) normal mode $\mu$ on site $n$ is thus obtained as the ratio of mode frequency shift and defect size in the limit of small shifts,
\begin{equation}\label{eq:weight}
W_{\mu n} = \frac{\Delta \Omega_{\mu|n}}{\Delta\omega_n} \;\to\; \frac{d\Omega_{\mu|n}}{d\omega_n}\bigg|_{\omega_n=0}.
\end{equation}

To image the normal modes experimentally, we thus need to insert a lattice defect of known size $\Delta\omega_n$ at the location of any given site $n$, and to measure the shift $\Delta\Omega_{\mu|n}$ in the normal-mode frequency resulting from the defect insertion. This is achieved as follows.
We create the defect with a dielectric probe consisting of a $2.2\times 2.2\,\text{mm}^2$ piece of sapphire. The dielectric is mounted to a cryogenic three-axis positioning stage inside the dilution refrigerator, so we can scan the probe in-situ across the  lattice and bring it into close proximity with each targeted resonator. Presence of the dielectric probe locally increases the resonator's capacitance per unit length $c$ and, thus, reduces its frequency $\omega_r/2\pi = 1/2L\sqrt{lc}$ ($l$ denotes the inductance per unit length, and $L$ the length of the resonator. The magnitude of the defect is easily tuned by varying the height of the probe above the surface of the resonator.

\section{Scanning the Defect Probe across a Single Resonator\label{sec:singleres}}
To validate the imaging performance of the Scanning Defect Microscope, we first employ the method to a single straight coplanar-waveguide resonator. We scan the dielectric probe in the lateral direction along the length of the resonator as well as in the vertical direction, changing the probe height above the resonator center. Experimental results for the two types of scans are presented in Fig.\ \ref{YZscan}. We obtain the frequency shift $\Delta\omega_r/2\pi$ via homodyne transmission measurements for the $\lambda/2$ mode and find very good agreement between the observed frequency shifts and those predicted by finite-element simulations using the HFSS package.

Data and simulation for the lateral scan [Fig.\ \ref{YZscan}(a)] match the intuitive expectation that positioning the dielectric probe in a region of small (large) electric field produces a small (large)  frequency shift. We observe a maximum shift of $663\,$MHz  at the field anti-nodes. The residual shift of $50\,$MHz at the node  is explained by the finite probe size, naturally causing the dielectric to cover field regions to the left and right of the zero-field resonator midpoint. Overall, we conclude that the simple picture of locally increasing the effective dielectric constant of the resonator and, hence, downshifting the resonator frequency serves as a good model to understand the data. 

For the purpose of imaging photon occupancy inside the lattice, it is crucial to calibrate the Scanning Defect Microscope and determine the dependence of defect size (i.e., resonator frequency shift) on probe height $z$ above the resonator,  $\Delta \omega_r=\Delta\omega_r(z)$. Our calibration data is shown in Fig.\ \ref{YZscan}(b) along with the prediction from numerical HFSS simulations. Here, the only noteworthy deviations observed are restricted to the range below probe heights of $25\,\upmu$m where small tilts in the orientation of the probe dielectric as well as small offsets in  measured vs.\ actual probe height $z$ can affect the data. We have verified with HFSS simulations that probe tilts as small as $0.02^\circ$ can lead to deviations of the same magnitude as those observed for our data, see inset of Fig.\ \ref{YZscan}(b). Since imaging a larger photon lattice by Scanning Defect Microscopy only requires accurate calibration for small frequency shifts at large $z$, we remain relatively impervious to small tilts and height offsets.

For a single resonator,  the induced defect size can thus be calibrated to the vertical distance of the scanning probe from the resonator, and the Scanning Defect Microscope matches the predictions from detailed numerical simulations. 

\begin{figure*}
\centering
\includegraphics[width=0.8\textwidth]{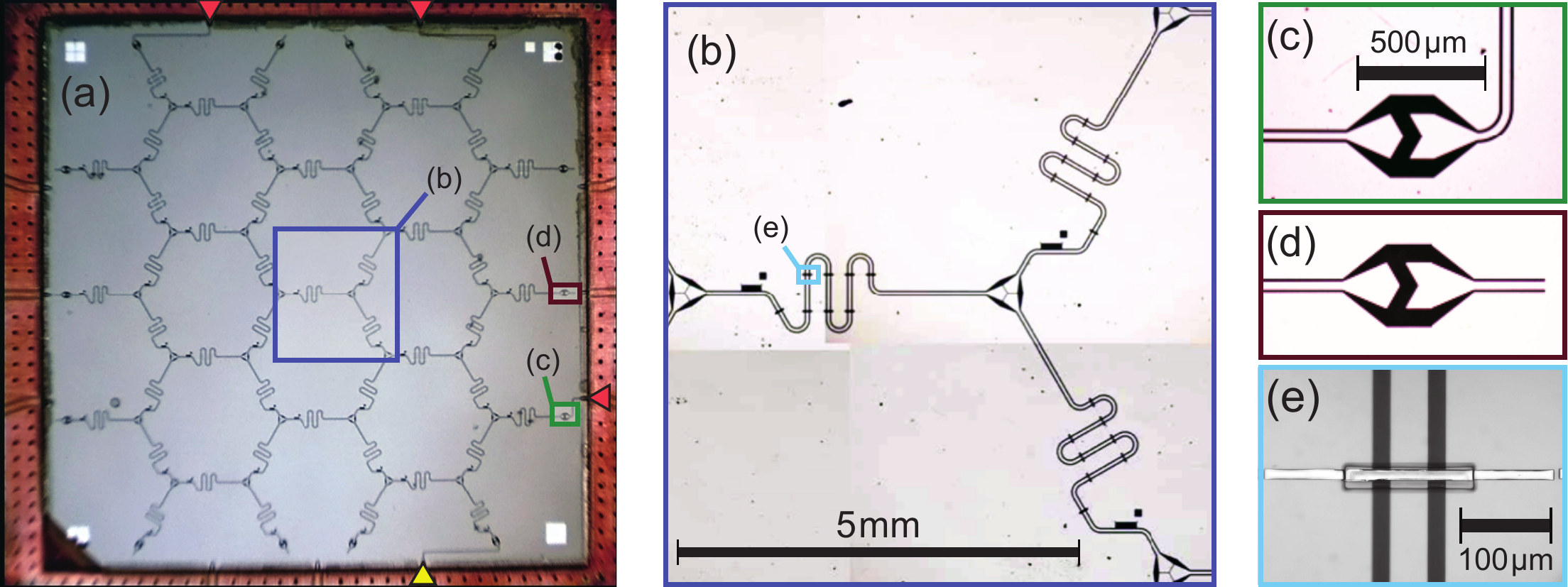}
\caption{Device picture of photon lattice, consisting of 49 coupled coplanar-waveguide resonators. (a) The resonators are arranged to form a Kagome lattice for photons (the dual of the honeycomb geometry visible when viewing the physical resonators). The device is equipped with a port used to drive the lattice with a coherent microwave tone (yellow triangle), and three possible ports to detect the transmitted signal (red triangles). (b) Resonators have a meandering section (to save space on the chip) and are coupled by three-way capacitors. Small capacitors (c), connect four of the edge resonators to ports and (d), terminate the remaining edge resonators by coupling to a high-frequency resonator stub. (e) Isolated ground planes within the lattice are connected with aluminum bridges evaporated on top of insulating  pads.}
\label{device}
\end{figure*}

\section{Scanning Defect Microscopy of photonic lattice}
After this successful validation of the defect probe performance for a single resonator, we now employ Scanning Defect Microscopy to a larger photon lattice and image its normal modes. Our experimental sample consists of an array of superconducting microwave resonators on a $32\times32\,\text{mm}^2$ chip,  forming a photonic Kagome lattice. 

Figure  \ref{device}(a) shows a picture of the 49 coupled resonators along with drive and measurement ports.  
The lattice bulk consists of resonators coupled via three-way capacitors [Fig.\ \ref{device}(b)]. At the edges of the lattice, we capacitively couple four resonators to input and output ports [Fig.\  \ref{device}(c)]. The input port [yellow triangle in Fig.\ \ref{device}(a)] is used to feed a coherent microwave signal with tunable drive frequency into the lattice. Three different output ports [red triangles in Fig.\ \ref{device}(a)] are used, one at a time, for transmission measurements. All remaining edge resonators are terminated by capacitive coupling to  high-frequency resonators [Fig.\ \ref{device}(d)]. In order to keep all edge resonators at the same frequency as the resonators in the bulk and not affect the normal modes of the lattice, the edge resonators are designed to have slightly altered lengths which aim to compensate for frequency offsets (see Appendix \ref{app:systematicDisorder} for details).

In the interior of the lattice, we ensure proper grounding of metal planes by interconnecting all planes via aluminum bridges evaporated onto a  bisbenzocyclobutene (BCB) spin-on glass [Fig.\ \ref{device}(e)]. For future realizations of Jaynes-Cummings lattices, our resonator design also includes a small cutout where a superconducting qubit can be inserted, visible in Fig.\ \ref{device}(b) as black rectangles close to one end of each resonator. All cutouts between resonator centerpins and ground planes are at the same relative distance from the nearest coupling capacitor and have the same orientation to allow future fabrication of the necessary Josephson junctions by double-angle evaporation.

\begin{figure*}
\centering
\includegraphics[width=0.85\textwidth]{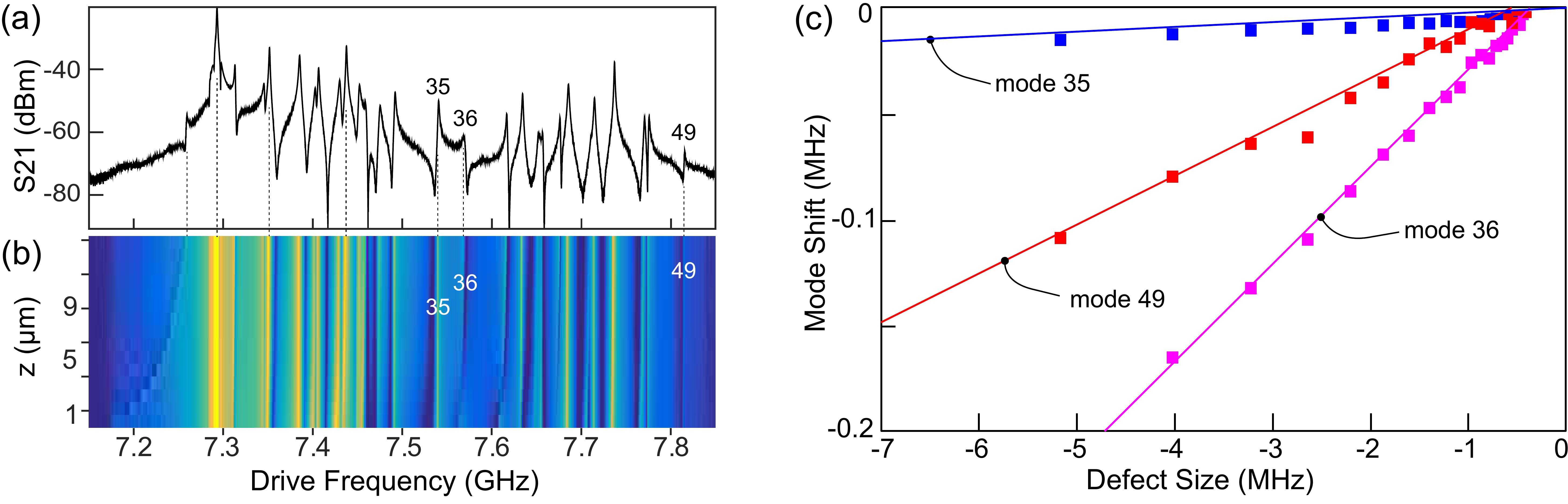}
\caption{Transmission and mode-shift data from Scanning Defect Microscopy. (a) The transmission spectrum of the Kagome lattice exhibits a multitude of resonances corresponding to normal modes which can be excited and detected through the selected input and output ports. (b) Positioning the scanning probe above one resonator and changing the vertical probe distance $z$ produces shifts in mode frequencies, here seen as ``bending'' of  maxima in the measured transmission S21 (color-coded), especially for small probe distances. (c) Calibration allows conversion of the probe height $z$ into the corresponding defect size $\Delta\omega_n$ of site $n$. For small defect size,  the normal-mode shifts $\Delta\Omega_{\mu|n}$ are linear in the $\Delta\omega_n$. Normal-mode weights are directly determined from the slope of each curve.}
\label{modeDefect}
\end{figure*}

The transmission spectrum for fully withdrawn scanning probe, measured across a select pair of input and output ports, shows the characteristic resonances associated with the unperturbed normal modes of the lattice [Fig.\ \ref{modeDefect}(a)].  Similar to previous findings for a 12-site Kagome lattice \cite{Underwood2012}, we note the presence of a faint resonance at the lowest end of the frequency spectrum. We attribute this resonance to the presence of the localized states characteristic of the Kagome lattice. Each of these degenerate states has alternating amplitudes limited to one hexagon of the Kagome lattice. (In the thermodynamic limit, it is these states that form the well-known flat band of the Kagome lattice.) While localization would ideally render these modes undetectable when driving and probing edge resonators,  slight amounts of disorder in resonance frequencies or hopping strengths weakens the localization and makes these modes visible in transmission.
 
To construct a complete map of mode weights for the lattice with Scanning Defect Microscopy, we follow the procedure from Section \ref{sec:model} and measure the mode frequency shifts induced by a small defect size on each lattice site. 
A single step in scanning the lattice consists of lateral positioning of the probe above one lattice site of the array, followed by a series of transmission measurements for varying probe heights. The necessary movements of the defect probe are performed in-situ, i.e., within the dilution unit. However, since our probe positioners have a limited motion range of $20\,$mm in both $x$ and $y$ direction, a single scan could not traverse the whole chip. Consequently, several cooldowns were necessary to span the larger range of the full lattice.

At each lattice site, we first center the probe over the resonator, then move it vertically down until contact is reached, and finally retract the probe vertically away from the surface (Details of the calibration for lateral positioning  are discussed in Appendix \ref{app:positionCal}.)
When in contact, 
the probe is approximately at the height of the BCB supported bridges, i.e., about $4\,\upmu$m above the resonator. (We will refer to this probe position as probe height $z$$=$$0$.) We retract the probe stepwise from contact, increasing $z$ 
up to a height of approximately $300\,\upmu$m, and perform transmission measurements for each probe position. Representative data  from our measurements with the probe above one particular site are shown in Figure \ref{modeDefect}(b) for the case of relatively small probe heights. When in such close proximity to the surface, the probe induces frequency shifts of normal modes that can exceed $100\,$MHz. For probe heights $z\ge 200\,\upmu$m, mode shifts are much smaller and, as expected, asymptotically approach zero.

To obtain the frequency shift $\Delta\Omega_{\mu|n}$ of normal mode $\mu$ due to a defect of size $\Delta\omega_n$ on site $n$, we track the resonance frequency of by fitting Lorentzians to the relevant part of the transmission spectra. In addition, we must convert the chosen vertical probe position $z$ into the corresponding defect size $\Delta\omega_n(z)$. This conversion task is non-trivial since the calibration from Section \ref{sec:singleres} for a straight resonator does not carry over quantitatively: resonators in our lattice sample include a meandering portion and feature multiple distinct orientations with respect to the square dielectric probe. From the simple picture of a piece of dielectric above a resonator, one expects that the defect size asymptotically obeys a simple scaling law $\Delta\omega_n = \gamma_n\omega_r/z^2$ for large probe distance $z$. However, the proportionality constant $\gamma_n$ may depend on the specific geometry and orientation of the resonator relative to the probe. 
For the conversion from vertical probe distance $z$ to defect size $\Delta\omega_n$, we have therefore performed finite-element simulations for the different types of lattice sites (see Appendix \ref{app:defectCalibration} for details of the defect calibration).

For the resulting defect sizes $\Delta\omega_n/2\pi$ with magnitude below $10\,$MHz, we observe that normal-mode shifts depend linearly on the defect size [Figs.\ \ref{modeDefect}(c)]. We attribute the observed offsets for modes 36 and 49, which would indicate a vanishing mode shift for non-zero defect size, to deviations of the experimental data from a strict $1/z^2$ scaling of defect size with probe height and resulting imperfections in defect size calibration for large $z$.

Recalling that the normal-mode weight of mode $\mu$ on site $n$ is given by
$W_{\mu n} = d\Omega_{\mu|n}/d\omega_n$ evaluated in the limit of $\Delta\omega_n=0$ [Eq.\ \eqref{eq:weight}], we see that the imaging of normal modes only relies on the data in the linear regime, i.e., the limit of small defect sizes. The normal-mode weights are directly obtained from the slopes of linear fits to the normal-mode shift data, such as shown for one particular resonator in Fig.\ \ref{modeDefect}(c).

\begin{figure*}
\centering
\includegraphics[width=1\textwidth]{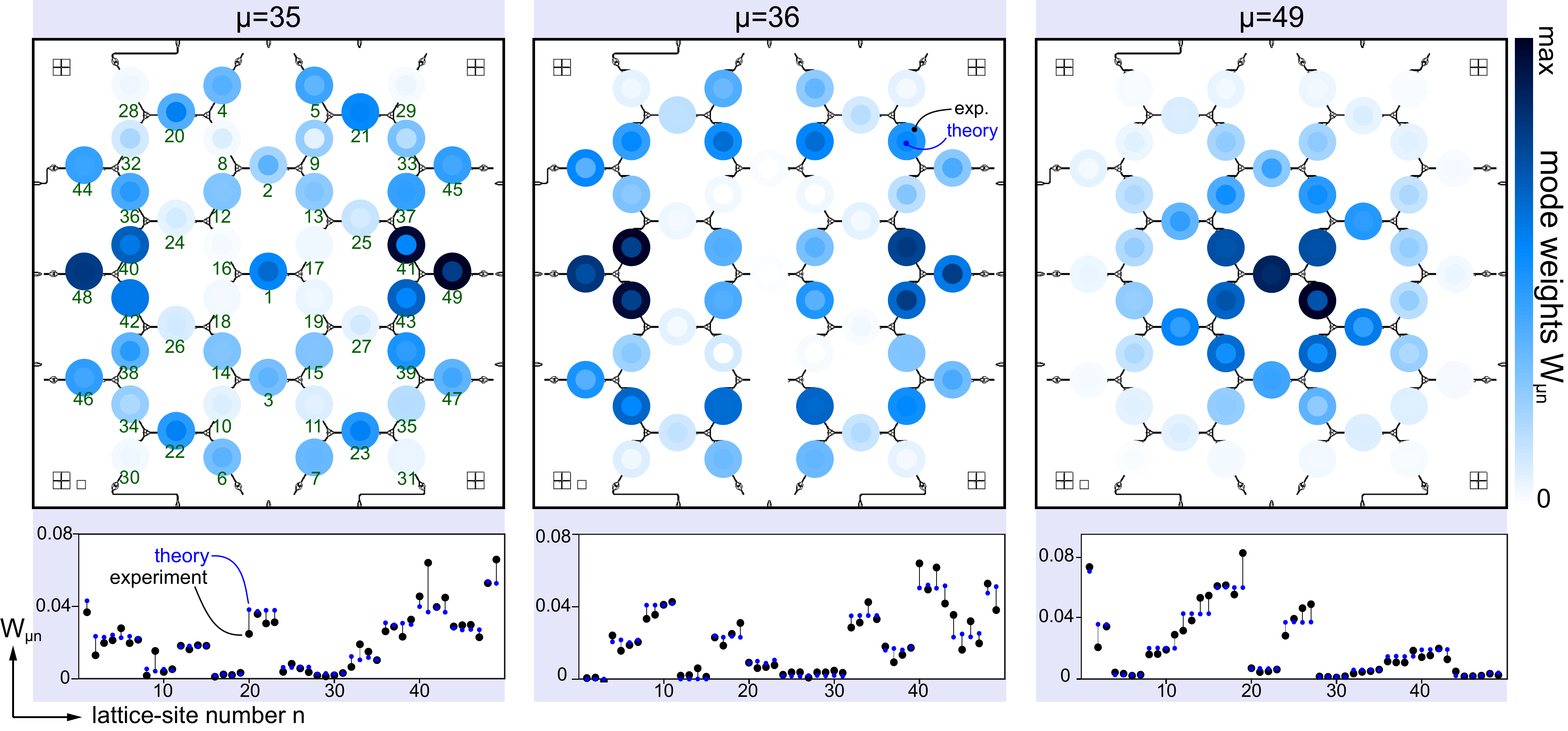}
\caption{Normal-mode weights for three select modes, $\mu=35,\,36$ and $49$. For each lattice site $n$, the mode weight $W_{\mu n}=|(n|\mu)|^2$ is depicted an overlay of two color-coded disks. Inner disks represent the theory prediction for the weights, the outer disks the weights determined experimentally by Scanning Defect Microscopy. The bottom panels show deviations in normal-mode weights as lines connecting experimental data (black dots) and theory prediction (blue dots).
Comparison shows good agreement with noticeable deviations in only a few sites in each case.
\label{fig:weights}
}
\end{figure*}

\section{Normal-mode images}
We present the results from Scanning Defect Microscopy of our sample in Fig.\ \ref{fig:weights}. For the normal-mode imaging, we have selected three modes which are well-separated and easily identifiable: the modes $\mu=35$ and $36$ fall into a sparse region of the spectrum, in the frequency range that would contain the Dirac point in the case of an infinite Kagome lattice; the third imaged mode, $\mu=49$, is the highest-frequency mode in our spectrum [Fig.\ \ref{modeDefect}(a--b)].

For each of the three modes, we perform an overall normalization step and then visualize the normal-mode weights $W_{\mu n}$ as color-coded disks centered on the 49 individual sites of the lattice [Fig.\ \ref{fig:weights}, top panels]. For easy comparison with theory, each disk is overlaid by a second disk of reduced size, displaying the theory prediction for the corresponding weight. Our theory calculation of mode weights accounts for a small amount of systematic disorder in frequencies among the different categories of resonators, as revealed by HFSS simulations. The mode images readily reveal that weights respect, to good approximation, the mirror symmetry along the vertical and horizontal axes crossing through the center site $n=1$. (The $\mathsf{D}_6$ symmetry of the full Kagome lattice is reduced to $\mathsf{D}_2$ due to finite size and geometry of our sample.) Overall, our mode images show good qualitative agreement with the normal-mode weights predicted by theory.

A quantity which is fairly insensitive to deviations in normal-mode weights from isolated outliers, but appropriately captures the overall qualitative agreement, is the fidelity 
\begin{equation}
\mathcal{F}_\mu=\sum_n (W_{\mu n}^\text{exp}\,W_{\mu n}^\text{th})^{1/2}.
\end{equation}
Assuming that the experimental mode amplitudes
$(n|\mu_\text{exp})=\pm (W_{\mu n}^\text{exp})^{1/2}$
carry the sign expected from theory (here just signs because mode amplitudes can be chosen real-valued), the fidelity reduces to the ordinary state overlap: 
\begin{equation}
\mathcal{F}_\mu=\sum_n(\mu_\text{exp}|n)(n|\mu_\text{th})=(\mu_\text{exp}|\mu_\text{th}). 
\end{equation}
Table \ref{table:compare} presents the fidelities achieved in our experiment. For all three modes the fidelities are close to 1, confirming good overall agreement.

For more detailed quantitative comparison between experimental data and theory not relying on perception of small color differences, we provide an alternative representation of the same normal-mode data in the bottom panels of Fig.\ \ref{fig:weights}. In these graphs, vertical line segments connect the predicted weights (blue circles) with the experimental weights (black circles), such that deviations between theory and experiment are easily read off as the lengths of the vertical line segments. The plots readily show that larger deviations are not specific to particular lattice sites. While we cannot rigorously rule out lattice disorder as a factor contributing to the observed deviations,  previous experience indicates that such lattice disorder may be expected to be small \cite{Underwood2012}. It is more likely that deviations are primarily owed to the difficulty of precise calibration of defect sizes for the large lattice, a procedure which relies on a simple scaling model and HFSS simulations within a restricted range of probe heights.

\begin{table}
	\caption{\label{table:compare}%
		Quantitative comparison between theoretical predictions and experimental results for normal-mode weights as obtained from Scanning Probe Microscopy of the 49-site Kagome lattice.}
	\begin{ruledtabular}
		\begin{tabular}{c c c}
			Mode   &    Fidelity  &   Normalized rms \\ 
			$\mu$ &  $\mathcal{F}_\mu=(\mu_\text{exp}|\mu_\text{th})$ & $\text{n-rms}_\mu$ (see text) \\[1mm]\hline
			35  & 0.990 & 12\% \\
			36  & 0.989 & 10\% \\
			49  & 0.993 & 9\% 
		\end{tabular}
	\end{ruledtabular}
\end{table}

To condense the deviations $\Delta W_{\mu n}=|W_{\mu n}^\text{exp}-W_{\mu n}^\text{th}|$ between experimental and theoretical normal-mode weights into a single figure of merit for each mode, we employ the normalized root-mean-square deviation (n-rms)
\begin{equation}
\text{n-rms}_\mu=\frac{(\textstyle\frac{1}{N}\sum_n\Delta W_{\mu n}^2)^{1/2}}{\max_n (W_{\mu n}^\text{th})-\min_n (W_{\mu n}^\text{th})}.
\end{equation}
As shown in Table \ref{table:compare}, n-rms values indicate that averaged deviations obtained in this way are of the order of 10\%. While not a high-precision measurement, Scanning Defect Microscopy thus provides us a with a detailed and quantitative image of normal-mode weights of the resonator lattice.

\section{Conclusions and Outlook}
We have introduced Scanning Defect Microscopy, a new tool for acquiring local information in photon lattices driven and measured at the lattice edge.  Transmission through a cavity lattice is measured as a dielectric probe is positioned over each cavity, with changes in transmission revealing the local weight of the normal mode.  Spatial maps of a single lattice resonator are used to validate the performance of the tool and to calibrate defect size.  In this paper, we have used Scanning Defect Microscopy to experimentally determine the normal-mode weights for chosen modes in a 49-site Kagome resonator lattice and observed good agreement with theory.  This technique will provide key insight into local properties of these lattices when interactions are strong, and an important tool for the study of non-equilibrium quantum phase transitions and quantum simulation.

\begin{acknowledgments}
The authors thank Darius Sadri, James Raftery, Guanyu Zhang, David Schuster and Hakan T\"ureci for fruitful discussions. This work
was supported by the National Science Foundation through Grants No. DMR-0953475 and No. PHY-1055993, by the Army Research Office under Contract No. W911NF-11-1-0086, and by the Packard Foundation. Additionally, D.L.U.\ acknowledges support by  National Science Foundation Graduate Research Fellowship No.\ DGE-1148900.
\end{acknowledgments}

\appendix
\section{Elimination of bulk/edge frequency differences
	\label{app:systematicDisorder}}
In the bulk of the Kagome lattice, resonators are coupled to four nearest neighbors (two on each side). By contrast, edge resonators are coupled to two bulk resonators on one end, and either a port (transmission line) or a far-detuned quarter-wavelength resonator on the other end. Without any compensatory measures, this would lead to a systematic frequency difference between bulk and edge resonators. In the design of our resonator lattice, we aim to eliminate this frequency difference by adjusting the length of edge resonators appropriately. To do so, we note that the mode frequencies of a transmission-line resonator with coupling capacitances $C_\alpha$ $(\alpha=L,R)$ at its two ends are determined by the transcendental equation \cite{Koch2010c}
\begin{equation}\label{taneq}
\tan (\bar{\omega}) = -\frac{(\chi_L + \chi_R)\bar{\omega}}{1-\chi_L\chi_R \bar{\omega}^2}.
\end{equation}
Here, $\bar\omega=\omega\sqrt{\ell c}L$ and $\chi_\alpha=C_\alpha/(cL)$ express the frequency and coupling capacitance in dimensionless form; $c$, $\ell$ and $L$ denote the capacitance per unit length, inductance per unit length, and total length of the resonator. According to Eq.\ \eqref{taneq}, changes in coupling capacitors indeed lead to a change in mode frequencies. We now determine the appropriate change in resonator length to compensate for this change, keeping in mind that both $\omega$ and $\chi_\alpha$ depend on this length.

To distinguish between bulk and edge resonators, we apply labels $s=b,e$ to frequencies, lengths and coupling capacitances ($c$ and $\ell$ remain the same for bulk and edge). The desired elimination of frequency differences, $\omega^e=\omega^b$, implies $\bar\omega^e = \bar\omega^b\cdot L^e/L^b$. Plugging this into Eq.\ \eqref{taneq} for the edge resonators,
\begin{equation}
\tan (\bar{\omega}^e) = -\frac{(\chi_L^e + \chi_R^e)\bar{\omega}^e}{1-\chi_L^e\chi_R^e (\bar{\omega}^e)^2},
\end{equation}
one finds that the modified edge-resonator length can be written in closed form as
\begin{equation}
L^e = -L^b \frac{1}{\bar{\omega}^b}\arctan\bigg[ \frac{(C_L^e+C_R^e)\bar{\omega}^b/cL^b}{1+C_L^eC_R^e(\bar{\omega}^b/cL^b)^2}\bigg]\simeq L^b\frac{\omega^e_\text{unadj.}}{\omega^b},
\end{equation} 
where $\omega^e_\text{unadj.}$ is the frequency of edge resonators if their length remains unadjusted. In the last step, we have discarded terms beyond the leading-order term in an expansion in $\Delta L/L^b$.

\section{Probe movement and position calibration \label{app:positionCal}}
We calibrate the lateral probe position on each lattice site by monitoring changes in the frequency spectrum as the probe is shifted in $x$ and $y$ direction. Moving the probe laterally across the lattice is implemented by retracting the probe from the surface to a height of $100 \upmu$m, moving it parallel to the surface by $100 \upmu$m and then lowering it again until good mechanical contact is reached. Each step therefore ends with the probe in mechanical contact with the surface of the lattice. We perform a transmission measurement after each step and record changes in the spectrum. The resonances presumably associated with quasi-localized modes are the most sensitive to the probe position, and sustain shifts as large as $200$MHz when the probe is centered over a lattice site.  For this reason, frequency shifts of these particular modes are convenient for inferring the probe position. (Position readings via potentiometric measurements of the integrated position encoders were not sufficient to directly obtain lattice coordinates. The encoders were used for coarse positioning and measurement of $z$ coordinates.)

\begin{figure}
\centering
\includegraphics[width=\columnwidth]{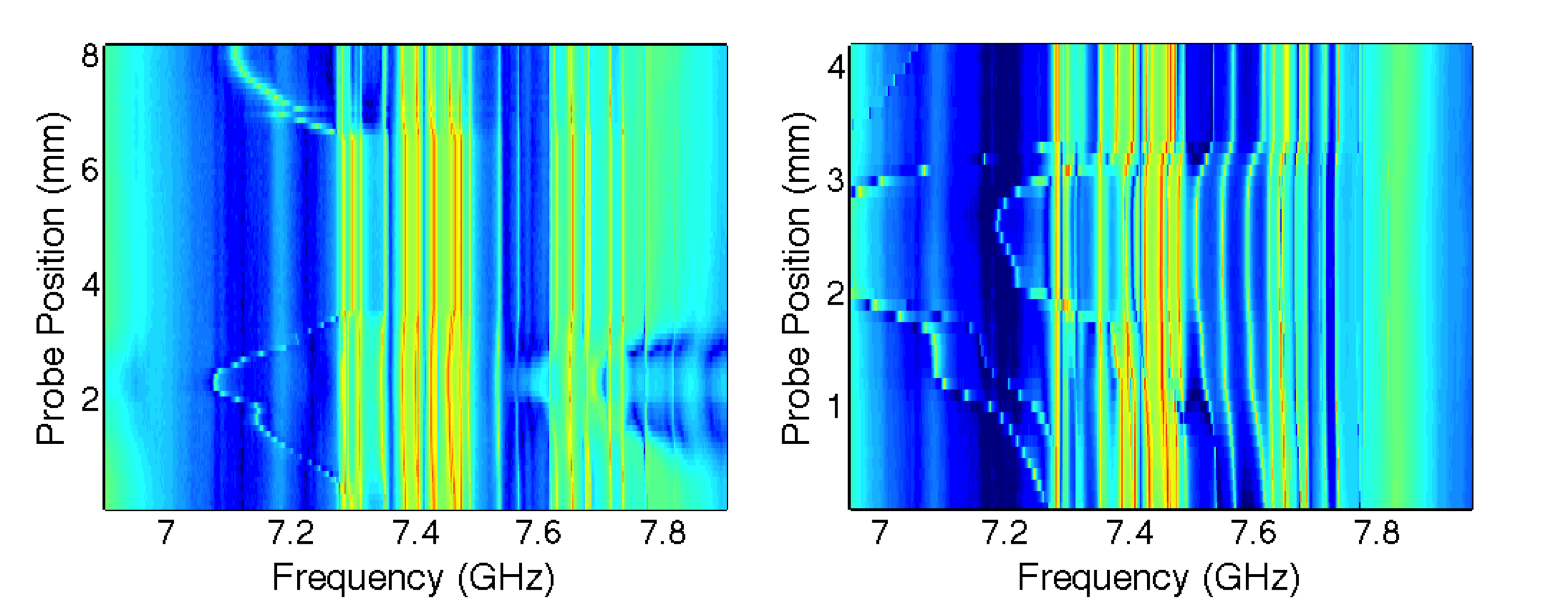}
\caption{Transmission spectra for different lateral probe positions. (a) Starting on a ground plane, the probe is moved across one lattice resonator, traverses the ground plane, and stops centered on a second lattice resonator. Significant shifts of the lowest-frequency mode result when the probe is centered on one resonator, no shifts occur when it is in contact with a ground plane. (b) Starting at the edge of a three-way coupling capacitor, the probe is moved across the capacitor, then off the capacitor and onto a lattice resonator. When the probe covers a three-way capacitor, three lattice sites are  perturbed simultaneously resulting in multiple modes shifting down in frequency.}
\label{PositionCalibration}
\end{figure}

The probe's $x$ and $y$ coordinates relative to a specific resonator are obtained by moving the probe off of the resonator onto the ground plane, and also by moving the probe onto a three-way coupler [Fig.\  \ref{PositionCalibration}]. Whenever the probe is positioned above the ground plane, no frequency shifts are observed; when the probe is positioned over a three-way coupler, multiple sites are perturbed and the frequency spectrum changes significantly. Together, this information is sufficient to achieve the needed lateral position calibration.

\section{Defect calibration \label{app:defectCalibration}}
By adjusting the vertical separation between a resonator and the probe, the defect size (i.e., the shift in resonance frequency of this particular resonator) can be tuned in a controlled manner. In the lattice geometry, this shift cannot be determined directly by transmission measurements and must hence be inferred by other means.

To calibrate the defect size to the probe height, $\Delta \omega_n=\Delta \omega_n(z)$ we have performed HFSS simulations for the bulk and edge resonators in the three occurring orientations and for multiple probe heights ranging from $40$ to $100\,\upmu$m. These data allow us to fix the proportionality constants in the scaling law $\Delta \omega_n=\Delta \omega_n(z)$ for the different categories of resonators.

Both the smallness of defect sizes at large probe heights and deviations from the simple $1/z^2$ scaling make the calibration challenging and are the dominant limiting factors of the precision in mode weight determination.

\end{document}